# Studies of Plasma Detachment Using a One Dimensional Model for Divertor Operation

R. A. Vesey [a] , D. E. Post [a,b], and G. Bateman [a]
[a] Princeton University Plasma Physics Laboratory, Princeton, NJ, USA
[b] ITER Joint Central Team, San Diego, CA, USA

## Abstract

To characterize the conditions required to reach advanced divertor regimes, a one-dimensional computational model has been developed based on a coordinate transformation to incorporate two-dimensional effects. This model includes transport of ions, two species each of atoms and molecules, momentum, and ion and electron energy both within and across the flux surfaces. Impurity radiation is calculated using a coronal equilibrium model which includes the effects of charge-exchange recombination. Numerical results indicate that impurity radiation acts to facilitate plasma detachment and enhances the power lost from the divertor channel in escaping neutral atoms by cooling the electrons and suppressing ionization. As divertor particle densities increase, cold and thermal molecules become increasingly important in cooling the plasma, with molecular densities dominating electron and atomic densities under some conditions.

## 1. Introduction

An understanding of the processes capable of sustaining advanced divertor operation (radiative or gas-target) is crucial for next-generation devices such as ITER[1] and TPX[2]. Current experiments have provided data on detached divertor plasmas which may be used to guide analytic and computational models. $D_2$ injection in DIII-D results in a factor of 3-5 reduction in the peak heat flux to the plate, preceded by the formation of a stable high density radiating region upstream [3]. This is accompanied by a sharp drop in $T_e$ and $n_e$ at the separatrix strike point, with little measurable change in the plasma parameters near the midplane and without the addition of non-intrinsic impurities. Borrass and Stangeby [4, 5] described the role of ion-neutral collisions in transferring ion momentum and energy to the nearby solid surface, accounting for the decrease in plasma pressure and ion flux observed in detached plasmas in JET. The effectiveness of ion-neutral collisions depends on sufficiently low values of $T_e$ in the divertor and a high probability that neutrals will escape the divertor channel after colliding with the ions. Radiation (due to hydrogen, intrinsic impurities, and injected impurities) and ion-neutral interactions are likely to play complementary roles in dissipating the power flowing into the scrape-off layer. The focus of this paper is to quantify, using simple geometry, the effects of intrinsic impurities and ion-neutral interactions.

## 2. Code description

A system of eight 1-D time-dependent fluid conservation equations is numerically solved using finite volume spatial discretization with upwind differencing applied to all convective terms. The difference equations are time-advanced implicitly. The system consists of particle balances for five species:

(1) ions,
(2) "cold" atoms (at 3 eV energy from Frank-Condon dissociation),
(3) "thermal" atoms (in equilibrium with ions at $T_i$),
(4) "cold" molecules (at 0.1 eV ≈ plate temperature), and
(5) "thermal" molecules (in equilibrium with ions at $T_i$);

(6) conservation of momentum parallel to the magnetic field for the combined thermal species (ions, thermal atoms, and thermal molecules) which are assumed to acquire the same parallel velocity, and two separate energy balances,

(7) thermal ion and atomic species (ions, thermal atoms, and thermal molecules), and
(8) plasma electrons.

Features of the physical model include:

(1) Transformation of both parallel and cross-field terms into a single spatial coordinate is performed by assuming the shape of plasma

contours [6] and is shown schematically in Figure 1. The key effect is to represent in a 1-D model the enhanced penetration of neutrals (atoms and molecules) entering from the side of the divertor channel ("recirculating" particles).

(2) The source of recirculating particles is the integrated flux of neutral particles which escape the divertor channel without first colliding with an ion. A simple escape probability (based on the divertor channel width and the mean-free-path) is applied whenever a thermal atom is created (typically by charge-exchange between a cold atom and an ion). The interactions of the recirculating particles are added as explicit source/sink terms to the system of conservation equations.

(3) The rates for atomic processes including electron-impact ionization, recombination, and hydrogen radiation energy loss per ionization are calculated using a full collisional-radiative treatment [7, 8]. Atom-ion charge-exchange rate coefficients are taken from Janev et al. [9].

(4) Atom-ion and molecule-ion elastic scattering effects are included using a simple multi-step formulation, with rate coefficients provided by Bachmann & Belitz [10, 11] . These effects provide sources of thermal atoms and thermal molecules due to upscattering of cold particles, and also enhance the cross-field diffusion of particles, momentum, and energy.

(5) The ratio of impurity density to hydrogenic nuclear density is assumed to be fixed and spatially uniform. The radiation rates are obtained from a coronal equilibrium model [12, 13] which accounts for the effect of charge-exchange recombination between impurity ions and hydrogen neutrals [14]. However, the impurity charge-exchange recombination does not self-consistently affect the ion and atom densities in the simulation.

(6) The modeled molecular processes include:

(a) molecular ionization followed immediately by dissociation:

$$H_2(v=0) + e^- \Rightarrow H_2^+(v) + 2e^-$$
$$H_2^+(v) + e^- \Rightarrow e^- + H^+ + H(1s)$$

(b) direct molecular dissociation by electron-impact:

$$H_2(v=0)+e^- \Rightarrow e^- + H(1s) + H(1s)$$

The molecular reactions are the standard set used in most divertor modeling codes[15], with reaction rates taken from Janev et al. [9].

## 3. Results for DIII-D parameters with carbon impurity

The simulations discussed in this section involve typical DIII-D parameters (connection length along magnetic field=26 m, major radius=1.7 m, scrape-off layer width=0.01 m, total parallel energy flux = 170 MW/m$^2$, corresponding to approximately 1.8 MW into the outer divertor leg). The plasma ion species is deuterium. (The terms "H" and "$H_2$" are used generically to refer to hydrogenic species regardless of isotope.) The plate recycling coefficient is set to 1.0. Uniform carbon impurity fractions of 1% and 3% (with respect to the total density of hydrogen nuclei) are assumed, which when extrapolated to the core correspond to Z-effective values of 1.28 and 1.76, respectively. The coordinate transformation assumes a value of 0.2 for tan(α) (see Figure 1). Below are the results of a scan of $N_h$, the total number of hydrogenic particles in the system (defined as the line integral of the density along the field line). The total number of particles consists of hydrogen ions, atoms, and molecules (x 2) and is the unique independent variable once the geometry, input power, and impurity fraction have been specified.

Figure 2 plots the densities at the divertor plate of the various particle species as a function of $N_h$, for the 1% carbon system. The beginning of plasma detachment is indicated by the vertical line, where detachment is defined to occur when the ionization/ radiation peak is separated from the divertor plate. Once the plasma detaches, an initial increase in the plate electron density is observed, but this saturates and begins to decline as neutral species begin to dominate. The molecular species densities (both "cold" plate-emitted molecules at 0.1 eV and "thermal" molecules in equilibrium with the ions) steadily increase once the plasma detaches. Eventually, the molecular species dominate all others near the plate. Also, for both the 1% and 3% carbon systems, the midplane electron density is relatively insensitive to $N_h$ once the plasma detaches. These results indicate that once the plasma detaches, the further addition of particles to the system results in a rapid increase in the molecular densities near the plate, with little change in the electron density at the midplane or plate.

The total radiated power (due to hydrogen, carbon impurity, and bremsstrahlung) is plotted in Figure 3 as a function of $N_h$. In this plot, the solid curves include the effects of charge-exchange recombination between the carbon impurity ions and neutral hydrogen in calculating the coronal equilibrium emission rates. The approximate onset of plasma detachment is indicated by the boxed points. The maximum radiated power is achieved once the plasma is fully detached, when the peak in the radiated power profile is well separated from the divertor plate. After full detachment, the slow rise as $N_h$ increases is due to the increase in the density of carbon impurity (for a fixed concentration) coupled with the relatively constant value of $n_e$ at the plate seen in Figure 2. For comparison, several simulations were performed in which the effect of charge-exchange recombination was ignored; these cases are indicated by the open symbols in Figure 3. For the fully-detached cases, charge-exchange recombination effects increase the total radiated power by about 20% for both the 1% and 3% carbon cases.

Power may also be lost from the scrape-off layer by neutral atoms which escape the divertor channel without colliding with the plasma ions or electrons. This loss mechanism is maximized when thermal atoms are created (by charge-exchange or recombination processes) in a region in which $T_i > T_e$. In this model, the local escape probability for thermal atoms increases as the ionization cross-section decreases. Therefore impurity radiation, which primarily acts to decrease the electron temperature (thereby decreasing the electron-impact ionization cross-section), is expected to enhance the power carried by thermal atoms which escape the divertor channel to the side walls. Figure 4 shows the total escaping neutral power versus $N_h$. The rapid rise in escaping neutral power at 3-5x$10^{18}$ total particles is explained by the sharp increase at the same time of the electron and cold molecule densities seen in Figure 2. This leads to an increase in the dissociation rate of cold molecules to cold atoms, which then undergo charge-exchange with the plasma ions and possibly escape to the side walls. The more gradual increase in the escaping neutral power above 6x$10^{18}$ total particles is due to the fact that once the plasma detaches, the addition of more particles to the system simply increases the length of the low $T_e$ region in front of the plate, expanding the region from which neutrals are likely to escape to the side walls. The escaping neutral power with 3% carbon exhibits a stronger dependence on the number of particles than that with 1% carbon because proportionately more carbon impurity is being introduced per hydrogen particle, thus radiating more electron energy and extending the neutral loss region.

Figure 5 shows the total power reaching the plate and the detachment distance as functions of $N_h$, for 1% and 3% carbon. The total power includes the heat reaching the plate in the ions, electrons, thermal atoms, and thermal molecules, as well as the 13.6 eV recombination potential energy carried by each hydrogen ion. (At the highest densities for both 1% and 3% carbon, the total power is approximately 50% heat content and 50% recombination potential energy.) The model assumes that all radiated power and escaping neutral power is lost to the sides of the divertor channel, which will certainly be invalid for very short detachment distances. Nevertheless, the total power reaching the divertor plate drops sharply as the region of strong ionization/radiation forms and then separates from the plate, followed by a more gradual decrease as the detachment distance increases. Also, the detachment distance in these steady-state simulations is a controllable parameter, given by the input power, impurity concentration, and total number of particles in the system.

## 4. Conclusions

Steady-state simulations have been performed to identify the behavior of plasma detachment, radiation, and escaping neutral power as a function of the total number of particles in the system. Radiation by impurities such as carbon at low concentrations (1-3%) facilitates plasma detachment for typical DIII-D divertor parameters. Charge-exchange recombination of the carbon impurity typically increases the total radiated power by 20% over cases in which the dependence of emissivity on neutral density is ignored. As $T_e$ at the plate decreases, molecular effects become important in further cooling the plasma, with molecular densities dominating ion and atom densities for a sufficiently high number of total particles in the system. At the same time, by cooling the electrons, impurity radiation allows atom-ion charge-exchange to compete with ionization, increasing the power and momentum dissipated by escaping atoms. Both radiated power and escaping neutral power increase dramatically as the plasma detaches, after which both gradually increase as more particles are added to the system. Although these simulations sought to quantify radiation and ion-neutral interactions using reasonable parameters for DIII-D, future work will focus on particular experimental results from DIII-D and other machines as more complete data become available.

Also to be investigated in future work is negative ion formation from vibrationally-excited molecules as a potentially important neutral source mechanism which can dominate three-body recombination at sufficiently low electron temperatures [7]. Further research is needed to estimate the actual fraction of molecules which are expected to be vibrationally-excited under divertor plasma conditions, and to gain better understanding of the energetics of the $H^-$ production process at low electron temperatures.

## Acknowledgment

The authors are grateful for discussions with M. Petravic, P-H. Rebut, and M. Rosenbluth. This research was supported in part by an appointment to the U.S. Department of Energy Distinguished Postdoctoral Research Program sponsored by the U.S. Department of Energy, Office of Science Education and Technical Information, and administered by the Oak Ridge Institute for Science and Education.

## References

1. G. Janeschitz, *Journal of Nuclear Materials* **(to appear)**, (1994).

2. D. Hill, Divertor Design for the Tokamak Physics Experiment, 11th International Conference on Plasma Surface Interactions in Controlled Fusion Devices (J. Nucl. Materials, Mito, Japan, 1994),

3. T. Petrie, et al., *Journal of Nuclear Materials* **196-198**, 849 (1992).

4. K. Borrass, P. Stangeby, *Proceedings of the 20th EPS Conference on Controlled Fusion and Plasma Physics* **II**, 763 (1993).

5. P. C. Stangeby, *Nuclear Fusion* **33**, 1695 (1993).

6. M. Watkins, P. Rebut, 19th European Conf. on Controlled Fusion and Plasma Physics Innsbruck, Austria, 1992), vol. 16C, Part II, pp. 731.

7. R. K. Janev, et al., *J. Nucl. Mater.* **121**, 10-16 (1984).

8. J. Weisheit, *Journal of Physics B (Atomic and Molecular Physics)* **8**, 2556 (1975).

9. R. K. Janev, W. D. Langer, K. E. Jr., D. E. Post, *Elementary Processes in Hydrogen-Helium Plasmas*. (Springer-Verlag, Gmbh, and Co., Berlin, 1987).

10. P. Bachmann, H. J. Belitz, Max Planck Institute for Plasma Physics, IPP 8/2, Elastic Proceses in Hydrogen-Helium Plasmas: Collision Data (1993).

11. D. Reiter, in ***Atomic and Plasma-Material Interaction Processes in Controlled Thermonuclear Fusion*** R. Janev, H. Drawin, Eds. (Elsevier, Amsterdam, 1993) pp. 243-266.

12. R. Hulse, *Nuclear Technology/Fusion* **3**, 259 (1983).

13. D. E. Post, R. V. Jensen, C. B. Tarter, W. H. Grasberger, W. A. Lokke, *Atomic Data and Nuclear Data Tables* **20**, 397-439 (1977).

14. R. A. Hulse, D. E. Post, D. R. Mikkelsen, *J. Phys. B* **13**, 3895-3907 (1980).

15. D. Heifetz, in *Physics of Plasma Wall Interactions* D. Post, R. Behrisch, Eds. (Plenum, New York, 1986), vol. 131, pp. 695-772.

## List of Figure Captions

Figure 1. Schematic of coordinate system and transformation.

Figure 2. Particle densities at the divertor plate vs. total number of hydrogen particles, for 1% carbon case. In this plot $n_e$ = electron density, $n_m^{cold}$ = cold molecular density, $n_m^{thermal}$ = thermal molecular density, and $n_o^{thermal}$ = thermal atom density.

Figure 3. Total radiated power vs. total number of hydrogen particles, for 1% and 3% carbon. The solid curves include the effects of charge-exchange recombination of carbon ions, while the open symbols ignore this effect.

Figure 4. Total power carried by escaping neutral atoms vs. total number of hydrogen particles, for 1% and 3% carbon.

Figure 5. Total power reaching the divertor plate and plasma detachment distance vs. total number of hydrogen particles, for 1% and 3% carbon. The total power includes the plasma heat content and the recombination potential energy.

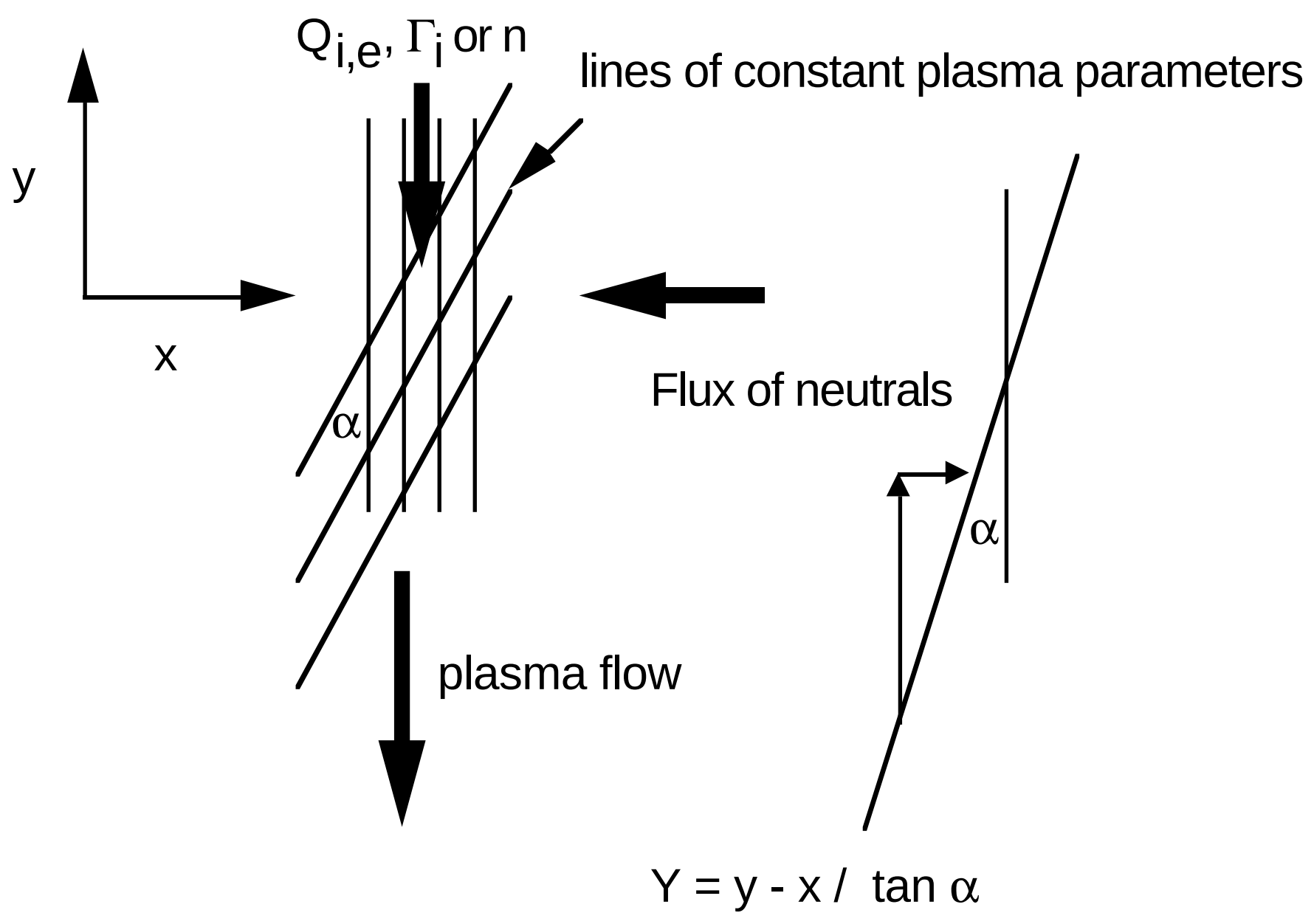

Figure 1. Schematic of coordinate system and transformation.

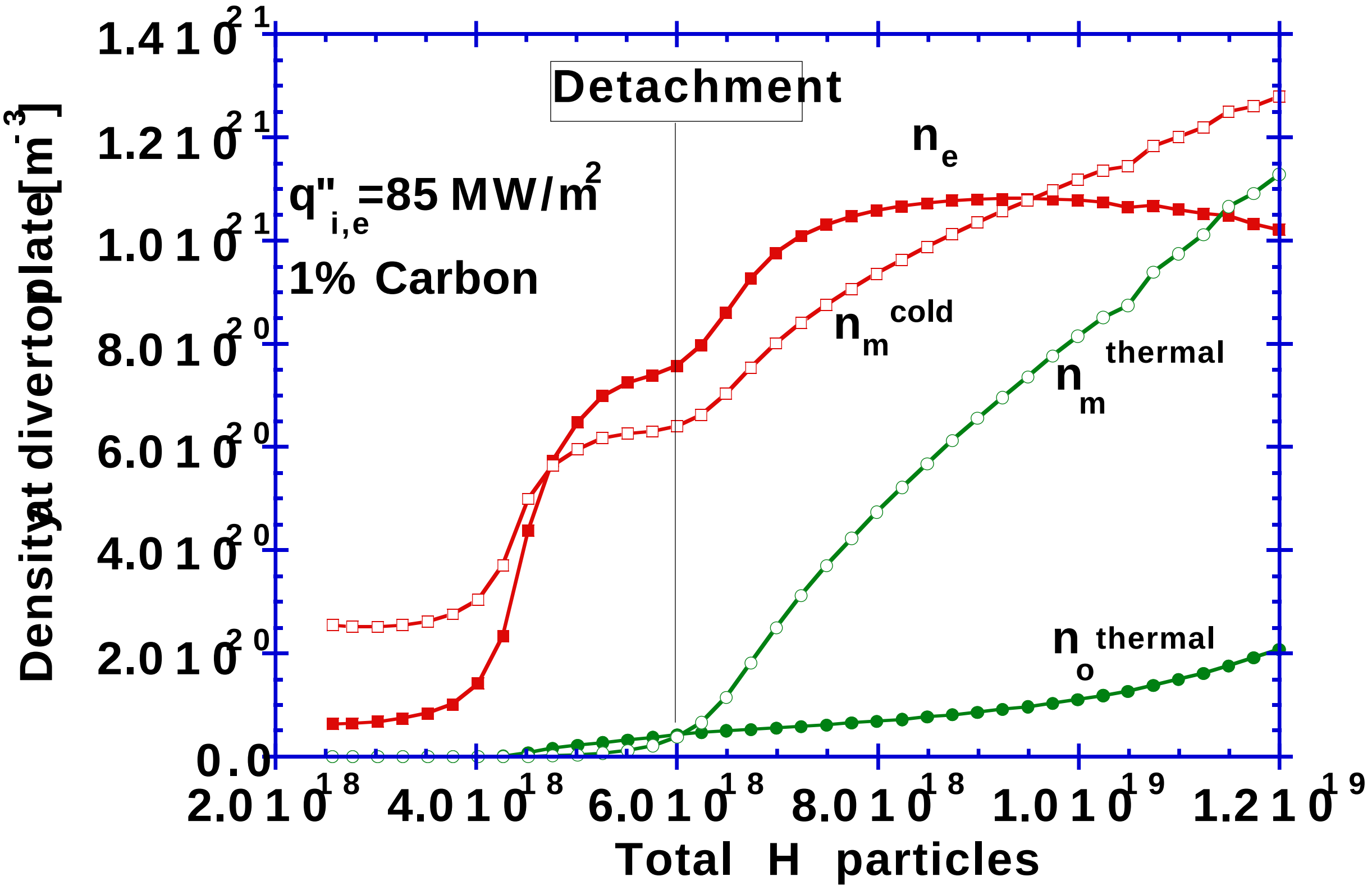

Figure 2. Particle densities at the divertor plate vs. total number of hydrogen particles, for 1% carbon case. In this plot $n_e$ = electron density, $n_m^{cold}$ = cold molecular density, $n_m^{thermal}$ = thermal molecular density, and $n_o^{thermal}$ = thermal atom density.

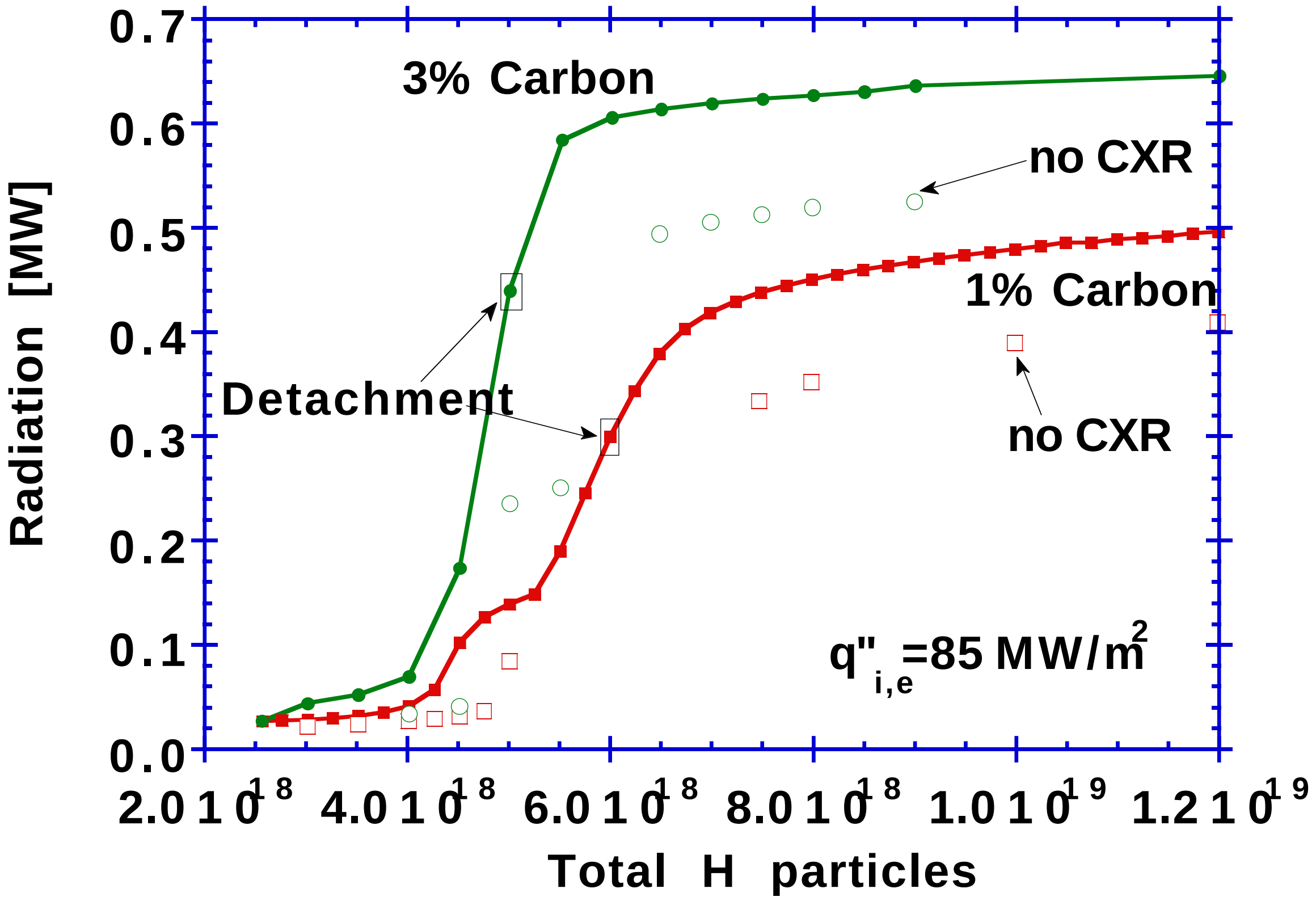

Figure 3. Total radiated power vs. total number of hydrogen particles, for 1% and 3% carbon. The solid curves include the effects of charge-exchange recombination of carbon ions, while the open symbols ignore this effect.

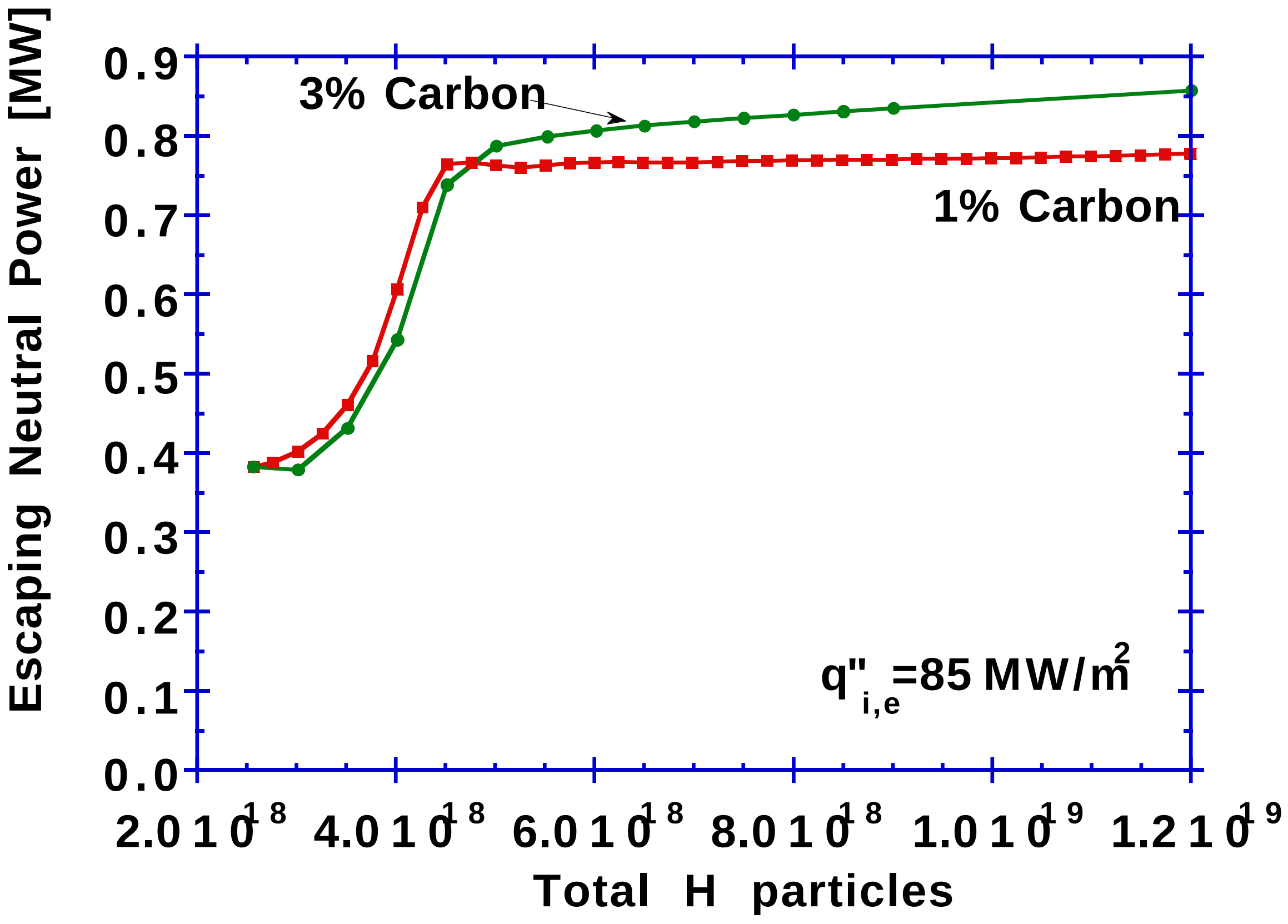

Figure 4. Total power carried by escaping neutral atoms vs. total number of hydrogen particles, for 1% and 3% carbon.

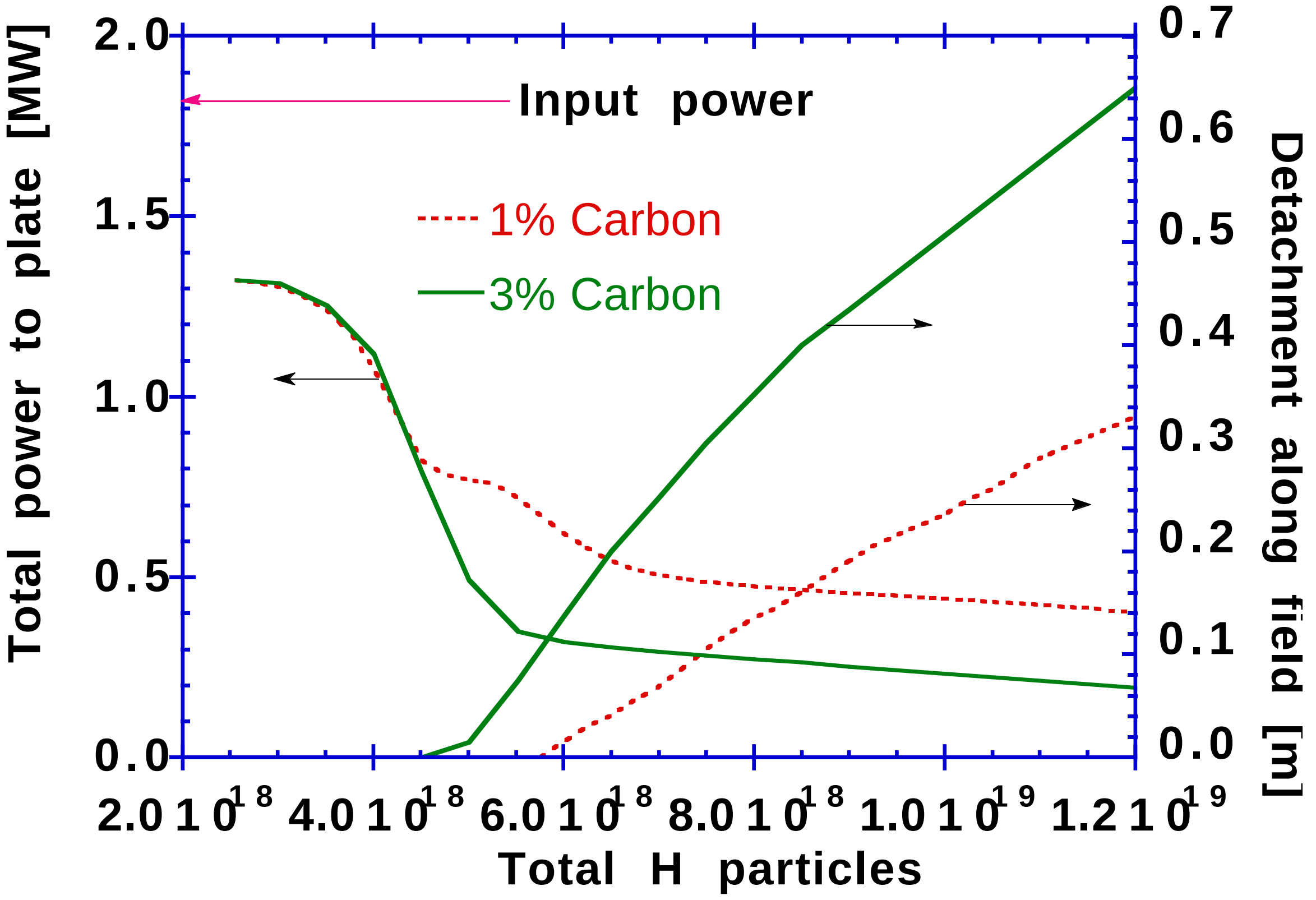

Figure 5. Total power reaching the divertor plate and plasma detachment distance vs. total number of hydrogen particles, for 1% and 3% carbon. The total power includes the plasma heat content and the recombination potential energy.